\documentclass[11pt]{article}

\usepackage{amsmath}
\usepackage{amssymb}
\usepackage{amsfonts}
\usepackage{gensymb}
\usepackage{slashed}
\usepackage{cite}
\usepackage{mathtools}

\usepackage{titlesec}
\usepackage{authblk}

\allowdisplaybreaks[2]

\newcommand{\ddx}{\mathrm{d}^4x\;}
\newcommand{\ddy}{\mathrm{d}^4y\;}
\newcommand{\ddp}{\mathrm{d}^4p\;}
\newcommand{\ddk}{\mathrm{d}^4k\;}
\newcommand{\drm}{\mathrm{d}}
\newcommand{\sgn}{\,\mathrm{sgn}}
\newcommand{\dels}{\Delta s}
\newcommand{\Tr}{\mathrm{Tr}}

\title{\textbf{Radiation reaction from quantum electrodynamics and its implications for the Unruh effect}}
\author[1]{Zolt\'an Tulip\'ant 
\thanks{e-mail: tulipantz@protonmail.com}
}
\affil[1]{ \small 
Institute for Theoretical Physics, E\"otv\"os Lor\'and University, H-1117 Budapest, 
\newline P\'azm\'any P\'eter s\'et\'any 1/A, Hungary
}

\begin{document}

\maketitle

\begin{abstract}
The Abraham--Lorentz--Dirac theory predicts vanishing radiation reaction for uniformly accelerated charges. However, since an 
accelerating observer should detect thermal radiation, the charge should be seen absorbing photons in the accelerated frame which, 
if nothing else occurs, would influence its motion. This means that either there is radiation reaction seen in an inertial frame 
or there should be an additional phenomenon seen in the accelerated frame countering the effect of absorption. In this paper I 
rederive the Abraham--Lorentz--Dirac force from quantum electrodynamics, then I study the case of a uniformly accelerated charge. 
I show that in the accelerated frame, in addition to the absorption of photons due to the Unruh effect there should also be 
stimulated emission. The net effect of these phenomena on the motion of the charge is found to be zero. 
\end{abstract}

\newpage

\section{Introduction}

Accelerated charges produce electromagnetic radiation that carries off energy according to the Larmor formula for radiated 
power\footnote{I will use natural units unless otherwise noted. In particular, these units are derived from SI units by setting 
$\hbar=c=\varepsilon_0=k_B=1$. Another popular choice is to set $4\pi\varepsilon_0=1$ instead which would make the constant 
multiplier in the Larmor formula $2e^2/3$.} \cite{Jackson:100964}
\begin{equation}
\label{Larmor}
P = -\frac{e^2}{6\pi} \ddot{x}^\mu\ddot{x}_\mu,
\end{equation}
where $e$ is the electric charge and $\ddot{x}^\mu\equiv\drm^2 x^\mu(s)/\drm s^2$ is the four-acceleration\footnote{I will use 
metric signature $(+ - - -)$ throughout the paper.}. If there is a loss of energy, there must also be a recoil force acting on 
the particle, called radiation reaction. For classical charges this can be described by the Abraham--Lorentz--Dirac (ALD) force 
\cite{Dirac:1938nz} 
\begin{equation}
\label{ald}
F^\mu_{ALD} = \frac{e^2}{6\pi}\Big[ \dddot{x}^\mu - \dot{x}^\mu (\dddot{x}^\nu\dot{x}_\nu) \Big].
\end{equation}
Since $x^\mu(s)$ gives the world-line of a massive particle, the four-velocity $\dot{x}^\mu(s)$ is time-like, in particular 
$\dot{x}^\mu\dot{x}_\mu=1$ and it is orthogonal to the four-acceleration, $\dot{x}^\mu\ddot{x}_\mu=0$. Using these relations we 
can recast Eq.~(\ref{ald}) in a more familiar form, 
\begin{equation}
F^\mu_{ALD} = \frac{e^2}{6\pi}\Big[ \dddot{x}^\mu + \dot{x}^\mu (\ddot{x}^\nu\ddot{x}_\nu) \Big].
\end{equation}
For a review of the ALD radiation reaction, see Ref.~\cite{Poisson:1999tv}. 

This description of radiation reaction has some peculiar features. For hyperbolic motion, which is the relativistic generalization 
of uniform acceleration, the ALD force vanishes. This might suggest the absence of electromagnetic radiation for eternal hyperbolic 
motion but this result has been shown to be non-paradoxical in Refs.~\cite{Fulton,Boulware}. In the case of bremsstrahlung, 
acceleration happens in a finite time interval and it is the difference of the fields supported by the accelerated and the 
inertial charge that is released as radiation. Perhaps the easiest way to cope with this is realizing that there can be no clear 
distinction made between radiation and other field configurations. 

In Refs.~\cite{Fulling,Davies:1975,Unruh1} it has been shown that a uniformly accelerated observer detects thermal radiation of 
temperature $T_U=a/2\pi$, where $a=\sqrt{-\ddot{x}^\mu\ddot{x}_\mu}$ is the magnitude of acceleration. This phenomenon became known 
as the Unruh effect. In the case of an accelerated detector with multiple levels of internal energy, like the Unruh-DeWitt detector, 
the effect of absorbing photons from the thermal background seen in the accelerated frame can be explained in an inertial frame by 
the emission of photons \cite{UnruhWald}. Such an emission, which is referred to as Unruh radiation, can cause recoil. For a 
uniformly accelerated charge that does not have excited states, however, there is no Unruh radiation and the ALD theory predicts 
vanishing radiation reaction. In the accelerated frame, nonetheless, the charge should be seen absorbing quanta from the thermal 
background which may cause the charge to recoil. Thus, either the act of absorption observed in the accelerated frame should be 
countered by another phenomenon or the ALD formula gives the wrong result. However, if there would be radiation reaction for 
uniformly accelerated charges, or equivalently, the Unruh effect would cause recoil, that would violate the equivalence principle 
since we could distinguish between an inertial charge and a charge in free fall. 

A problematic trait of the ALD radiation reaction is that it admits unphysical solutions, like preacceleration, i.e.~the 
particle starts accelerating even before any external force is applied. This preacceleration has a characteristic time which 
in SI units is $t_0=\mu_0 e^2/(6\pi m c)$, where $m$ is the mass of the particle. Such violation of causality might not be a serious 
problem. For an electron $t_0\sim10^{-24}s$, hence, in the case of subatomic particles this unphysical behavior appears in 
situations when quantum corrections should be applied \cite{Rohrlich2}. Furthermore, Eq.~(\ref{ald}) is valid for point-like 
charges and it has been shown that when taking finite-size effects into account, there is no violation of causality \cite{Yaghjian}. 
This provides some clues as to under what circumstances does the ALD theory give reliable results. 

In order to assess what is the full classical (or semiclassical) description of radiation reaction we should turn to quantum 
electrodynamics (QED) which is the most precise theory of the electromagnetic interaction to date. Its classical limit correctly 
reproduces Maxwell's equations. Although Dirac's original derivation of the ALD force appearing in Ref.~\cite{Dirac:1938nz} was 
based on Maxwell's equations, effort has been made to reconstruct the classical radiation reaction from QED 
\cite{Higuchi:2002qc,Higuchi:2004pr,Higuchi:2005gh,Higuchi:2005an,Ilderton:2013tb,Polonyi:2017xdb}. 

In the following, I will rederive the classical radiation reaction for a charged particle. I start in section \ref{sect:integrating}  
by constructing an effective action from QED by integrating out the electromagnetic field of the charge. Then, in section 
\ref{sect:classical} I obtain the classical equation of motion in an inertial frame. Since in an accelerated frame the photon 
propagator contains the contribution of a thermal background, I will use the previously mentioned results in section 
\ref{sect:unruh} to determine how the Unruh effect influences the charged particle. In section \ref{sect:bremsstrahlung} I consider 
a charge that undergoes uniform acceleration only for a finite amount of time and point out that the results of section 
\ref{sect:unruh} apply in this case as well. Finally, in section \ref{sect:ehrenfest} I also derive a relation analogous to 
Ehrenfest's theorem that establishes a connection between expectation values.

\section{Integrating over the gauge field}
\label{sect:integrating}

In the path integral approach to quantizing field theories we construct a generating functional using the action of the classical 
field theory. The action for the electromagnetic field coupled to a charged field or particle characterized by a $\{q_j\}_{j=1}^M$ 
set of variables is 
\begin{equation}
\label{action}
S[\{q_j\}_{j=1}^M,A] = K[\{q_j\}_{j=1}^M] + \int{\ddx \left[ -J^\mu A_\mu - \frac{1}{4}F_{\mu\nu}F^{\mu\nu}\right]},
\end{equation}
where $F_{\mu\nu}=\partial_\mu A_\nu - \partial_\nu A_\mu$ denotes the field strength for the gauge field $A_\mu$. The functional 
$K[\{q_j\}_{j=1}^M]$ is the kinetic term and $J^\mu$ is the current density for the field or particle. If it is a fermionic field described by the spinor $\psi$ that is coupled to the electromagnetic field, these terms are 
\begin{equation}
K_{f}[\bar{\psi},\psi] = \int{\ddx \bar{\psi}(i\slashed{\partial}-m)\psi, \qquad 
J_{f}^\mu = e\bar{\psi}\gamma^{\mu}\psi }. 
\end{equation}

Let us separate the electromagnetic field into the field of the charge and an external field described by $A_c^\mu$ and $A_e^\mu$ 
respectively. Then the vector potential in Eq.~(\ref{action}) is replaced by $A_c^\mu+A_e^\mu$. Assuming that the external gauge 
field can be treated classically and that the charge has little to no effect on its source, we can keep $A_e^\mu$ fixed and 
neglect its contribution in the kinetic term $-F^{\mu\nu}F_{\mu\nu}/4$. Thus, the only contribution of the external field will be 
in the interaction term of the charge. This can be seen by expanding the kinetic term of the electromagnetic field, 
\begin{equation}
\label{kineticrewrite}
-\frac{1}{4}\int{\ddx\ F^{\mu\nu}F_{\mu\nu}} = -\frac{1}{4}\int{\ddx F_c^{\mu\nu}F_{c,\,\mu\nu}} - \frac{1}{4}\int{\ddx F_e^{\mu\nu}F_{e,\,\mu\nu}} - \int{\ddx \partial^\mu A_c^\nu F_{e,\,\mu\nu}}, 
\end{equation}
where $F_{c,\,\mu\nu}$ and $F_{e,\,\mu\nu}$ denote the field strength of the field of the charge and the external gauge field 
respectively. The third term can be rewritten using integration by parts as 
\begin{equation}
\label{discardable}
-\int{\ddx \partial_\mu A_{c,\,\nu} F_e^{\mu\nu}} = \int{\ddx A_{c,\,\nu} \partial_\mu F_e^{\mu\nu}}.
\end{equation}
Since the external field is handled classically, $\partial_\mu F_e^{\mu\nu}=J_e^\nu$ is the current density associated with 
the source of the external field. The assumption that the radiation field should have little to no effect on the source of 
the external field means that $J_e^\mu$ is peaked far away from the region where $A_c^\mu$ gives a significant contribution and 
approaches zero everywhere else. Hence, the overlap between $A_c^\mu$ and $J_e^\mu$ is negligible and the integral of 
Eq.~(\ref{discardable}) can be discarded. The second term on the right-hand side of Eq.~(\ref{kineticrewrite}) can be dropped as 
well since it gives only a constant contribution to the action.

To deduce how radiation affects the motion of its source we can use the path integral formalism and construct an effective 
action for the charge by considering the contribution of every $A_c^\mu$ field configuration. I define the effective action 
$S_{eff.}[\{q_j\}_{j=1}^M]$ that arises from integrating over the field of the charge as 
\begin{align}
\label{eff}
\mathcal{N}\exp\Bigg[i\;S_{eff.}[\{q_j\}_{j=1}^M]\Bigg] = 
\int{ \mathcal{D}A_c }\; \exp\Bigg[& i\,K[\{q_j\}_{j=1}^M] \;+\; i\,\int{ \ddx } \;\Bigg\{ - J_\mu A_e^\mu - J_{\mu}A_c^{\mu} \nonumber \\ 
& - \frac{1}{4}F_{c,\,\mu\nu}F_c^{\mu\nu} - \frac{1}{2\lambda}(\partial_\mu A_c^{\mu})^2 \Bigg\} \Bigg],
\end{align}
where the integral on the right-hand side is a functional integral over the configurations of the gauge field $A_c^\mu$ and 
the normalization constant $\mathcal{N}$ does not depend on any field variable. The additional term 
$\frac{1}{2\lambda}(\partial_\mu A_c^\mu)^2$ takes care of gauge fixing.

In order to obtain the effective action, the kinetic term of the electromagnetic field is cast into a more manageable form,   
\begin{equation}
\int{\ddx \left(-\frac{1}{4} F_{c,\,\mu\nu} F_c^{\mu\nu} - \frac{1}{2\lambda}(\partial_\mu A_c^{\mu})^2\right) 
= 
\int{\ddx \frac{1}{2} A_c^\mu\left[g_{\mu\nu}\partial^2-\left(1-\frac{1}{\lambda}\right)\partial_\mu\partial_\nu\right]A_c^\nu}}.
\end{equation}
Performing the functional integral of Eq.~(\ref{eff}) over the gauge field we get 
\begin{equation}
\label{eff1}
S_{eff.}[\{q_j\}_{j=1}^M] = K[\{q_j\}_{j=1}^M]\; - \;\int{\ddx J_\mu(x) A_e^\mu(x)} 
\; \pm \;\int{\ddx \int{\ddy \Bigg\{\frac{1}{2}J^\mu(x)\Delta_{\mu\nu}(x,y)J^\nu(y) \Bigg\}}}. 
\end{equation}
The sign of the self-interaction term is linked to the regularization of the photon propagator $\Delta_{\mu\nu}(x,y)$ that is 
the solution of the equation 
\begin{equation}
\label{photon}
\left[g_{\mu\nu}\partial^2-\left(1-\frac{1}{\lambda}\right)\partial_{\mu}\partial_{\nu}\right]\Delta^{\nu\kappa}(x,y) = 
-\delta_\mu^\kappa\;\delta^{(4)}(x-y),
\end{equation}
where the derivatives are taken with respect to $x^\mu$. The solutions to this equation can be obtained, for example, by using 
Fourier-transformation. The four possible solutions for the Fourier-transform of the Green's function are given by 
\begin{align}
\label{propfourier}
\widetilde{\Delta}^{ret}_{\mu\nu}(p) &= \frac{1}{(p_0+i0)^2 - |\boldsymbol{p}|^2}\left[ g_{\mu\nu} - (1-\lambda)\frac{p_\mu p_\nu}{p^2} \right], \nonumber \\
\widetilde{\Delta}^{adv}_{\mu\nu}(p) &= \frac{1}{(p_0-i0)^2 - |\boldsymbol{p}|^2}\left[ g_{\mu\nu} - (1-\lambda)\frac{p_\mu p_\nu}{p^2} \right], \nonumber \\
\widetilde{\Delta}^F_{\mu\nu}(p) &= \frac{1}{p^2+i0}\left[ g_{\mu\nu} - (1-\lambda)\frac{p_\mu p_\nu}{p^2} \right], \nonumber \\
\widetilde{\Delta}^{F*}_{\mu\nu}(p) &= \frac{1}{p^2-i0}\left[ g_{\mu\nu} - (1-\lambda)\frac{p_\mu p_\nu}{p^2} \right], 
\end{align}
where $|\boldsymbol{p}|=\sqrt{p_1^2+p_2^2+p_3^2}$. In the following I will use the Feynman gauge $\lambda=1$. Expressing the 
propagator in position space, $\widetilde{\Delta}^{ret}_{\mu\nu}(p)$ and $\widetilde{\Delta}^{adv}_{\mu\nu}$ yield the retarded 
and advanced propagators respectively,
\begin{align}
\label{prop1}
\Delta^{ret}_{\mu\nu}(x,y) &= -g_{\mu\nu}\frac{1}{2\pi}\Theta(x^0-y^0)\delta((x-y)^2), \nonumber \\
\Delta^{adv}_{\mu\nu}(x,y) &= -g_{\mu\nu}\frac{1}{2\pi}\Theta(y^0-x^0)\delta((x-y)^2),
\end{align}
while $\widetilde{\Delta}^F_{\mu\nu}(p)$ leads to the Feynman propagator 
\begin{equation}
\label{prop2}
\Delta^F_{\mu\nu}(x,y) = g_{\mu\nu}\frac{i}{4\pi^2}\frac{1}{(x-y)^2 - i0}, 
\end{equation}
and $\Delta^{F*}_{\mu\nu}(x,y)$ is its complex conjugate. From these expressions we can construct the Wightman functions as 
\begin{equation}
\Delta^{\pm}_{\mu\nu}(x,y) = \pm\frac{1}{2}\Big(\Delta^{ret}_{\mu\nu}(x,y)-\Delta^{adv}_{\mu\nu}(x,y)\Big) 
+ \frac{1}{2}\Big(\Delta^F_{\mu\nu}(x,y)-\Delta^{F*}_{\mu\nu}(x,y)\Big), 
\end{equation}
which describe the propagation of positive- and negative-frequency modes respectively as evidenced by their Fourier-transforms, 
\begin{equation}
\widetilde{\Delta}^{\pm}_{\mu\nu}(p) = -2\pi i g_{\mu\nu}\Theta(\pm p_0)\delta(p^2).
\end{equation}
The calculation of the different propagators is detailed in appendix A. Radiation reaction arises from the real part of the 
positive-frequency Wightman function, which is the same as half of the difference of the retarded and advanced propagators, 
\begin{align}
\label{retadv}
\mathrm{Re}\Delta^+_{\mu\nu}(x,y) = \frac{1}{2}\Big(\Delta^{ret}_{\mu\nu}(x,y) - \Delta^{adv}_{\mu\nu}(x,y)\Big) 
= -g_{\mu\nu}\frac{1}{4\pi}\sgn(x^0-y^0)\delta((x-y)^2), 
\end{align}
since this expression violates time reversal symmetry. However, substituting this expression into Eq.~(\ref{eff1}) cancels the 
self-interaction term. In fact, this problem is rooted in a more general deficiency of the ordinary Lagrangian formalism, namely 
its inability to take dissipative effects into account. This shortcoming can be remedied by utilizing Schwinger's closed time path 
(CTP) formalism \cite{Schwinger:1961} which is suitable for describing systems where the forward and backward evolutions are 
dictated by different dynamics. This has been successfully used before to obtain radiation reaction \cite{Polonyi:2014239}. Here, 
the CTP effective action is defined as 
\begin{equation}
\label{SCTP}
S^{CTP}_{eff.}[\{q_j,\bar{q}_j\}_{j=1}^N] = 
S^{(1)}[\{q_j\}_{j=1}^N] - S^{(1)}[\{\bar{q}_j\}_{j=1}^N] + S^{(2)}[\{q_j,\bar{q}_j\}_{j=1}^N], 
\end{equation}
where the kinetic term and the interaction with the external field are contained in 
\begin{align}
&S^{(1)}[\{q_j\}_{j=1}^N] = K[\{q_j\}_{j=1}^N] - \int{\ddx J^\mu(x)A_{e,\,\mu}(x)}, \nonumber \\
&S^{(1)}[\{\bar{q}_j\}_{j=1}^N] = K[\{\bar{q}_j\}_{j=1}^N] - \int{\ddx \bar{J}^\mu(x)A_{e,\,\mu}(x)},
\end{align}
and the self-interaction is given by 
\begin{align}
S^{(2)}[\{q_j,\bar{q}_j\}_{j=1}^N] = -\frac{1}{2}\int\ddx\int\ddy
\begin{pmatrix}
J^\mu(x) & \bar{J}^\mu(x)
\end{pmatrix}
\begin{pmatrix}
\Delta^F_{\mu\nu}(x,y) & -\Delta^+_{\mu\nu}(x,y) \\
-\Delta^-_{\mu\nu}(x,y) & -\Delta^{F*}_{\mu\nu}(x,y)
\end{pmatrix}
\begin{pmatrix}
J^\nu(y) \\ \bar{J}^\nu(y)
\end{pmatrix},
\end{align}
following Ref.~\cite{Polonyi:2014239}. The variables $q_j$ and $\bar{q}_j$ and their corresponding current densities, $J^\mu(x)$ and 
$\bar{J}^\mu(x)$ describe the field or particle according to the forward and the inverse of the backward evolution respectively.  Exploiting the relation $\Delta^-_{\mu\nu}(x,y) = \Delta^+_{\mu\nu}(y,x)$, the self-interaction term can be rewritten as 
\begin{align}
S^{(2)}[\{q_j,\bar{q}_j\}_{j=1}^N] = -\frac{1}{2}\int\ddx\int\ddy\Bigg[ 
&J^\mu(x)J^\nu(y)\Delta^F_{\mu\nu}(x,y) - \bar{J}^\mu(x)\bar{J}^\nu(y)\Delta^{F*}_{\mu\nu}(x,y) \nonumber \\
&- 2J^\mu(x)\bar{J}^\nu(y)\Delta^+_{\mu\nu}(x,y) \Bigg], 
\end{align}
where the first two terms are self-energy corrections and the third term is responsible for radiation reaction.

\section{Classical equation of motion for a point charge}
\label{sect:classical}

Considering a point charge with mass $m$ coupled to the electromagnetic field, the kinetic term and the current density takes 
the form 
\begin{align}
\label{ppart}
&K_p[x,N] = \int{\drm s\; \left[-\frac{m}{2} \Big(N^{-1}(s)\dot{x}^\mu(s)\dot{x}_\mu(s)+N(s)\Big) \right] }, \nonumber \\ 
&K_p[\bar{x},\bar{N}] = \int{\drm s\; \left[-\frac{m}{2} \Big(\bar{N}^{-1}(s)\dot{\bar{x}}^\mu(s)\dot{\bar{x}}_\mu(s)+\bar{N}(s)\Big) \right] }, \nonumber \\ 
&J_p^\mu(x) = e\int{\drm s\; \dot{x}^\mu(s) \delta^{(4)}(x-x(s)) }, \nonumber \\
&\bar{J}_p^\mu(x) = e\int{\drm s\; \dot{\bar{x}}^\mu(s) \delta^{(4)}(x-\bar{x}(s)) },
\end{align}
where $x^\mu(s)$ describes the world-line of the particle following the forward evolution dynamics, while $\bar{x}^\mu(s)$ 
represents the world-line of the particle governed by the inverse of backward evolution dynamics. For dissipative systems, the two 
dynamics are different. $K_p[x,N]$ and $K_p[\bar{x},\bar{N}]$ are functionals of the world-line while $J_p^\mu(x)$ and 
$\bar{J}_p^\mu(x)$ are functions of space-time coordinates. $N(s)$ is the lapse function \cite{Hartle:1986yu} which plays the role 
of a Lagrange multiplier. The lapse function can be used for gauge fixing, as the action is invariant under reparametrizations of 
the world-line. The first term in the CTP effective action of Eq.~(\ref{SCTP}) becomes 
\begin{equation}
S^{(1)}_p[x,N]=\int{\drm s\;\left[-\frac{m}{2}\Big(N^{-1}(s)\dot{x}_\mu(s)\dot{x}^\mu(s)+N(s)\Big) 
- e A_{e\,\mu}(x(s))\dot{x}^\mu(s)\right]}, 
\end{equation}
and the second term, $S^{(1)}_p[\bar{x},\bar{N}]$ has the same functional form. The self-interaction term takes the form 
\begin{align}
S^{(2)}_p[x,\bar{x}] = -\frac{e^2}{2}\int\drm s\, \int\drm s'\, \Bigg\{&
\dot{x}^\mu(s) \dot{x}^\nu(s')\Delta^F_{\mu\nu}(x(s),x(s')) \nonumber \\
&-\dot{\bar{x}}^\mu(s) \dot{\bar{x}}^\nu(s')\Delta^{F*}_{\mu\nu}(\bar{x}(s),\bar{x}(s')) \nonumber \\
&+2\dot{x}^\mu(s) \dot{\bar{x}}^\nu(s')\Delta^+_{\mu\nu}(x(s),\bar{x}(s'))\Bigg\}. 
\end{align}
The full effective action is 
\begin{equation}
\label{Seff}
S^{CTP}_p[x,\bar{x},N,\bar{N}] = S^{(1)}_p[x,N] - S^{(1)}_p[\bar{x},\bar{N}] + S^{(2)}_p[x,\bar{x}].
\end{equation}
By using this effective action, we omit the creation of charged particle--antiparticle pairs and the effect of charged loops 
which places additional limitations on the validity of this theory besides those discussed in the previous section. 

The classical equation of motion can be obtained by first finding the extremum of the effective action, thus we need to compute 
its variation, that is, 
\begin{align}
\delta S^{CTP}_p[x] = \int\drm s\; \Bigg[&\frac{\delta S^{CTP}_p[x,\bar{x},N,\bar{N}]}{\delta x^\mu(s)} \,\delta x^\mu(s) 
+ \frac{\delta S^{CTP}_p[x,\bar{x},N,\bar{N}]}{\delta N(s)} \,\delta N(s) \nonumber \\
&+ \frac{\delta S^{CTP}_p[x,\bar{x},N,\bar{N}]}{\delta \bar{x}^\mu(s)} \,\delta \bar{x}^\mu(s)
+ \frac{\delta S^{CTP}_p[x,\bar{x},N,\bar{N}]}{\delta \bar{N}(s)} \,\delta \bar{N}(s) \Bigg].
\end{align}
The first term will provide the equation of motion for $x^\mu(s)$ and the second term takes care of the constraint on the 
four-velocity with $N(s)$ being interpreted as the rate of change of the proper time with respect to the world-line parameter 
$s$. The remaining terms produce the same equations for $\bar{x}^\mu(s)$ with the sign of the radiation reaction term altered. 
The functional derivative of the action with respect to $x^\mu(s)$ is 
\begin{align}
\label{xvar}
\frac{\delta}{\delta x^\mu(s)} S^{CTP}_p[x,\bar{x},N,\bar{N}] =& 
m \frac{\drm}{\drm s}\Big(N^{-1}(s)\dot{x}_\mu(s)\Big) - e F_{e,\,\mu\nu}\dot{x}^\nu(s) \nonumber \\
&+ e^2\int{\drm s'\;\dot{x}^\beta(s')\frac{\drm}{\drm s}\Delta^F_{\mu\beta}(x(s),x(s'))} \nonumber \\
&- e^2\int{\drm s'\;\dot{x}^\alpha(s)\dot{x}^\beta(s')\partial_\mu\Delta^F_{\alpha\beta}(x(s),x(s'))},
\nonumber \\
&- e^2\int{\drm s'\;\dot{\bar{x}}^\beta(s')\frac{\drm}{\drm s}\Delta^+_{\mu\beta}(x(s),\bar{x}(s'))} \nonumber \\
&+ e^2\int{\drm s'\;\dot{x}^\alpha(s)\dot{\bar{x}}^\beta(s')\partial_\mu\Delta^+_{\alpha\beta}(x(s),\bar{x}(s'))},
\end{align}
while the functional derivative with respect to the $N(s)$ lapse multiplier yields 
\begin{equation}
\label{lvar}
\frac{\delta}{\delta N(s)} S^{CTP}_p[x,\bar{x},N,\bar{N}] = 
\frac{m}{2}\Big(N^{-2}(s)\dot{x}_\mu(s)\dot{x}^\mu(s) - 1\Big).
\end{equation}
In Eq.~(\ref{xvar}), the value of the field strength tensor 
$F_{e,\,\mu\nu}=\partial_\mu A_{e,\,\nu}(x(s)) - \partial_\nu A_{e,\,\mu}(x(s))$ is taken along the world-line of the charge. For 
a stationary action Eq.~(\ref{lvar}) gives the constraint $\dot{x}^\mu(s)\dot{x}_\mu(s)=N^2(s)$ and Eq.~(\ref{xvar}) yields 
\begin{align}
\label{cEOM}
m\ddot{x}_\mu(s) - e F_{e,\,\mu\nu}\dot{x}^\nu(s) 
&+ e^2 \int \drm s'\;\dot{x}^\alpha(s) \dot{x}^\beta(s') 
\Big[ \partial_\alpha\Delta^F_{\mu\beta}(x(s),x(s')) - \partial_\mu\Delta^F_{\alpha\beta}(x(s),x(s')) \Big]
\nonumber \\
&- e^2 \int \drm s'\;\dot{x}^\alpha(s) \dot{x}^\beta(s') 
\Big[ \partial_\alpha\Delta^+_{\mu\beta}(x(s),x(s')) - \partial_\mu\Delta^+_{\alpha\beta}(x(s),x(s')) \Big]
= 0. 
\end{align}
Here, and from now on until otherwise noted, proper time will be used as the parameter $s$ to simplify equations. Furthermore, the 
classical equation of motion is obtained by setting $x^\mu(s)=\bar{x}^\mu(s)$ and as a consequence, the imaginary parts that come from 
the propagators cancel. Note that Eq.~(\ref{cEOM}) actually provides a semiclassical solution in the sense that the charge and the 
external field are treated classically while the field of the charge itself is treated quantum-mechanically. 

Substituting Eq.~(\ref{retadv}) in Eq.~(\ref{cEOM}) we get 
\begin{equation}
\label{aldprefin}
m_R\ddot{x}^\mu = e F_{e,\,\nu}^\mu\dot{x}^\nu + \frac{e^2}{6\pi}\Big[\dddot{x}^\mu - \dot{x}^\mu (\dddot{x}^\nu\dot{x}_\nu)\Big], 
\end{equation}
where the divergent self-energy term coming from the real part of the Feynman-propagator has been removed by defining 
\begin{equation}
m_R=m-\frac{e^2}{8\pi}\lim_{\epsilon\rightarrow0+}\frac{1}{\epsilon}, 
\end{equation}
which is the one-loop renormalized mass. The details of the calculation can be found in appendix B. The last term on the 
right-hand side of Eq.~(\ref{aldprefin}) is exactly the ALD force of Eq.~(\ref{ald}). 

As I have mentioned before, by using Eq.~(\ref{ppart}) we omit the creation of charged particle--antiparticle pairs and loop 
effects caused by charged particles. This means that if the charge is accelerated for proper time\footnote{Both $P$ in 
Eq.~(\ref{Larmor}) and $m_Rc^2$ are scalars and so we must divide the latter by the elapsed proper time.} $\Delta\tau$, the average 
power of radiation should be much smaller than $m_Rc^2/\Delta\tau$. Using Eq.~(\ref{Larmor}) and temporarily restoring SI units we 
get 
\begin{equation}
\label{constraint}
1\gg \frac{\mu_0 e^2}{6\pi m_R c^3} \overline{a^2} \Delta\tau \equiv t_0 \frac{\overline{a^2}\Delta\tau}{c^2},
\end{equation}
where $\overline{a^2}$ is the average squared magnitude of the acceleration. Thus, the circumstances under which preacceleration is 
predicted to occur lie outside the domain where this effective theory is reliable. The relation in Eq.~(\ref{constraint}) can be 
expressed in terms of the average squared change in four-velocity $\overline{\Delta v^2}$ as 
\begin{equation}
\frac{t_0}{\Delta\tau}\frac{\overline{\Delta v^2}}{c^2} \ll 1. 
\end{equation}

\section{Radiation reaction and the Unruh effect}
\label{sect:unruh}

For hyperbolic motion the squared four-acceleration is constant, i.e.~$\ddot{x}^\mu\ddot{x}_\mu=-a^2$. The world-line of a particle undergoing such motion can be expressed in an inertial coordinate frame as 
\begin{equation}
x^0(s)=\frac{1}{a}\sinh(as), \quad x^1(s)=\frac{1}{a}\cosh(as), \quad x^2(s)=0, \quad x^3(s)=0,
\end{equation}
where $s$ is the proper time and the initial conditions have been chosen such that $\dot{x}^1(0)=\dot{x}^2(0)=\dot{x}^3(0)=0$, 
$x^1(0)=1/a$ and $x^2(0)=x^3(0)=0$. The coordinates are chosen such that $\ddot{x}^\mu$ is in the $(x^0,x^1)$ plane. The invariant 
interval between two points of the world-line is 
\begin{equation}
(x-x')^2=\frac{4}{a^2}\sinh^2\left(\frac{a}{2}(s-s')\right).
\end{equation}
Substituting this into Eq.~(\ref{prop2}) we get the following expression for the Feynman propagator
\begin{equation}
\Delta^F_{\mu\nu}(x,x') = g_{\mu\nu}\frac{i}{4\pi^2}\frac{a^2}{4}\frac{1}{\sinh^{2}\left(\frac{a}{2}(s-s')\right) - i0},
\end{equation}
which can be also written following Ref.~\cite{Troost:1977} as 
\begin{align}
\label{RindlerFeynman}
\Delta^F_{\mu\nu}(x,x') &= g_{\mu\nu}\int{\frac{\ddk}{(2\pi)^4}e^{-ik_0(s-s')}\Bigg[\frac{1}{k^2 + i0} - \frac{2\pi i \delta(k^2)}{e^{|k_0|/T_U}-1}\Bigg]} \nonumber \\
&=g_{\mu\nu}\int{\frac{\ddk}{(2\pi)^4}e^{-ik_0(s-s')}\Bigg[ \mathcal{P}\frac{1}{k^2} 
- 2\pi i\delta(k^2)\Bigg\{\frac{1}{2} + \frac{1}{e^{|k_0|/T}-1} \Bigg\} \Bigg]},
\end{align}
where $\mathcal{P}$ denotes Cauchy principal value. This form makes it explicit that an observer undergoing hyperbolic motion 
is immersed in a thermal bath of photons with temperature $T_U=a/2\pi$ as seen from the accelerated frame. We can cast the 
positive-frequency Wightman function in a similar form, 
\begin{align}
\label{RindlerWightman}
\Delta^+_{\mu\nu}(x,x') = -2\pi i\,g_{\mu\nu}\int\frac{\ddk}{(2\pi)^4}e^{-ik_0(s-s')}\delta(k^2)\Bigg\{ 
&\Theta(k_0)\Bigg[1+\frac{1}{e^{|k_0|/T}-1}\Bigg] \nonumber \\ 
&+ \Theta(-k_0)\Bigg[\frac{1}{e^{|k_0|/T}-1}\Bigg] \Bigg\},
\end{align}
which clearly contains both positive and negative frequency contributions. This is due  to the fact that the notions of positive 
and negative energy are frame-dependent and what is seen as a positive-energy mode by an inertial observer is described as a 
mixture of positive- and negative-energy modes by the accelerated observer. Hence, $\Delta^+_{\mu\nu}(x,x')$ should be called the 
Minkowski-positive-frequency Wightman function. 

Using Eq.~(\ref{cEOM}) we can put the classical equation of motion in flat space-time in the form 
\begin{equation}
m_R\Big(\ddot{x}^\mu(s) + \Gamma^\mu_{\alpha\beta}\dot{x}^\alpha(s)\dot{x}^\beta(s)\Big) = 
e F^\mu_{e,\,\alpha}\dot{x}^\alpha(s) + F_{rr}^\mu,
\end{equation}
where $\Gamma^\mu_{\alpha\beta}$ are the Christoffel-symbols, $F_{rr}^\mu$ is the term that is responsible for radiation reaction 
and in an inertial frame it takes the form 
\begin{equation}
\label{Frr}
F_{rr}^\mu = - e^2 \int{\drm s' \dot{x}^\alpha(s)\dot{x}^\beta(s') 
\Bigg[\delta^\mu_\beta\partial_\alpha\Delta^{rr}(x,x') - g_{\alpha\beta}\partial^\mu\Delta^{rr}(x,x')\Bigg]}.
\end{equation}
The real part of the Feynman propagator is absorbed into the renormalized mass and its imaginary part is combined with 
the Minkowski-positive-frequency Wightman function as 
\begin{align}
\label{emab}
\Delta^{rr}(x,x') &= \Delta^+(x,x') - i\,\mathrm{Im}\Delta^F(x,x') \nonumber \\
&= -2\pi i\int\frac{\ddk}{(2\pi)^4}e^{-ik_0(s-s')}\delta(k^2)\Bigg\{
\Theta(k_0)\Bigg[1 + \frac{1}{e^{|k_0|/T}-1}\Bigg] + \Theta(-k_0)\Bigg[\frac{1}{e^{|k_0|/T}-1}\Bigg] \nonumber \\
&\hspace{160pt}-\Bigg[\frac{1}{2} + \frac{1}{e^{|k_0|/T}-1}\Bigg]\Bigg\}.
\end{align}
I factorized each propagator as $\Delta^F_{\mu\nu}(x,x')=g_{\mu\nu}\Delta^F(x,x')$ and 
$\Delta^+_{\mu\nu}(x,x')=g_{\mu\nu}\Delta^+(x,x')$. In order to interpret this expression let us consider a source minimally coupled to a bosonic field. The interaction Hamiltonian is just the linear combination of the creation and annihilation operators 
$\hat{a}_k^\dagger$ and $\hat{a}_k$ with coefficients $C_+(k)$ and $C_-(k)$, integrated over the phase space. The probability 
amplitude for emitting an additional particle into a background of $n$ identical particles with wave vector $k_\mu$ is 
\begin{equation}
\mathcal{A}_k(n\rightarrow n+1) 
= i\langle n+1|(C_-(k) \hat{a}_k + C_+(k) \hat{a}^\dagger_k)|n\rangle = i\sqrt{n+1} C_+(k).
\end{equation}
For $n=0$ we get that $iC_+(k)$ is simply equal to the amplitude $\mathcal{A}_k(0\rightarrow 1)$ of emitting a single particle 
into the vacuum. If the particles that constitute the background follow a Bose-Einstein distribution with temperature $T$ then 
to obtain the probability of emitting a particle with frequency $k_0=\omega>0$ we need to compute the sum of squared amplitudes 
weighted with $p_n=e^{-n\omega/T}\left(1-e^{-\omega/T}\right)$, which are just Boltzmann factors normalized such that 
$\sum_n p_n=1$. The probability of emission becomes 
\begin{align}
\sum_{n=0}^\infty p_n \;\int{\drm\Omega_k |\mathcal{A}_k(n\rightarrow n+1)|^2} 
&= \left(1-e^{-\omega/T}\right) \sum_{n=0}^\infty{(n+1)e^{-n\omega/T}}\;\int{\drm\Omega_k |\mathcal{A}_k(0\rightarrow 1)|^2} \nonumber \\ 
&= \left(1+\frac{1}{e^{\omega/T}-1}\right)\;\int{\drm\Omega_k|\mathcal{A}_k(0\rightarrow 1)|^2}, 
\end{align}
where integration over the spatial direction of $k_\mu$ has been carried out. We can obtain the probability of absorption 
similarly, 
\begin{equation}
\sum_{n=1}^\infty p_n \;\int{\drm\Omega_k |\mathcal{A}_k(n\rightarrow n-1)|^2} 
= \frac{1}{e^{\omega/T}-1}\;\int{\drm\Omega |\mathcal{A}_k(1\rightarrow 0)|^2}.
\end{equation}
Now it is clear that the first term in Eq.~(\ref{emab}) is due to emission into a thermal background while the absorption of 
quanta is taken into account via the second term. Although the propagators are associated with the vacuum to vacuum transition in 
the inertial frame, the definition of vacuum depends on the choice of coordinates and in an accelerated frame the Minkowski vacuum 
is seen as a thermal background. Hence, in the Rindler frame the propagator contains the contributions of the absorption of photons from the thermal background and apparently an emission process stimulated by the Unruh effect.

Note that although the expression in Eq.~(\ref{Frr}) is manifestly covariant and its indexed constituents behave as tensors under 
Lorentz-transformations, it is unsuitable for use in an arbitrary coordinate frame. The reason is that $\dot{x}^\mu(s)$ and 
$\dot{x}^\mu(s')$ are the tangent vectors of the world-line at different points in space-time and one of them must be subjected 
to parallel transport before any sensible comparison can take place. This problem can be circumvented by replacing $g_{\mu\nu}$ 
in the expression of the propagator with the bitensor $\mathfrak{P}_{\mu\nu}(x,x')$ for parallel transport between points $x$ and 
$x'$. This would require the calculation of products of the Christoffel-symbols $\Gamma^\mu_{\alpha\beta}$, making the calculation 
tedious. Alternatively, we can use Eq.~(\ref{Frr}) to express the integrand of $F_{rr}^\mu$ in terms of vectors independent of 
$s'$ before changing to a non-inertial frame, 
\begin{align}
F_{rr}^\mu &= -e^2\int{\drm s'\;\left[\dot{x}^\mu(s')\frac{\drm}{\drm s}\Delta^{rr}(x,x')-
\dot{x}^\nu(s)\dot{x}_\nu(s')\partial^\mu\Delta^{rr}(x,x')\right]} \nonumber \\
&= -e^2\int{\drm s'\;\left[\dot{x}^\mu(s')\frac{\drm}{\drm s}\Delta^{rr}(x,x') 
- \cosh(a(s-s'))\partial^\mu\Delta^{rr}(x,x')\right] }.
\end{align}
The derivative $\partial_\mu\Delta^{rr}(x,x')$ can be computed along the lines of Eq.~(\ref{deltarel}) of appendix B. This time, 
however, we need to express the partial derivative using the derivative with respect to $s$ instead of $s'$, thus obtaining 
\begin{equation}
\label{Frrinpro}
F_{rr}^\mu = -e^2\int{\drm s'\;\frac{\drm}{\drm s}\Delta^{rr}(x,x')\left[\dot{x}^\mu(s')
 - \cosh(a(s-s'))\frac{x^\mu(s)-x^\mu(s')}{\dot{x}_\nu(s)(x^\nu(s)-x^\nu(s'))}\right] }.
\end{equation}
At this point it is useful to introduce the Rindler coordinates $\xi^\mu$ which are related to the original Minkowski coordinates by 
\begin{equation}
x^0 = \xi^1 \sinh(a\xi^0), \quad x^1 = \xi^1 \cosh(a\xi^0), \quad x^2=\xi^2, \quad x^3=\xi^3,
\end{equation}
where $\xi^1\in\;]0,\infty[\;$ and $\xi^0,\xi^2,\xi^3\in\;]-\infty,\infty[\;$. These coordinates cover only the right Rindler 
wedge but that is sufficient for this discussion. In this frame the non-zero coordinates of the particle are 
\begin{equation}
\xi^0(s)=s, \qquad \xi^1(s)=1/a,
\end{equation} 
and the non-zero elements of the metric are 
\begin{equation}
g_{00} = (a\xi^1)^2,\qquad g_{11}=g_{22}=g_{33}=-1.
\end{equation}
The non-trivial relations between the different coordinate bases are 
\begin{align}
\tilde{t}_\mu &= \frac{1}{a\xi^1}\cosh(a\xi^0)\tilde{\eta}_\mu \;-\;\sinh(a\xi^0)\tilde{\zeta}_\mu, \nonumber \\  
\tilde{x}_\mu &= -\frac{1}{a\xi^1}\sinh(a\xi^0)\tilde{\eta}_\mu \;+\;\cosh(a\xi^0)\tilde{\zeta}_\mu, 
\end{align}
where $\tilde{t}_\mu$, $\tilde{x}_\mu$, $\tilde{\eta}_\mu$ and $\tilde{\zeta}_\mu$ are the basis vectors adapted for the coordinates 
$x^0$, $x^1$, $\xi^0$ and $\xi^1$ respectively. The non-zero Christoffel-symbols are 
\begin{equation}
\Gamma^{1}_{00}=a^2\xi^1,\qquad \Gamma^{0}_{01}=\Gamma^{0}_{10}=\frac{1}{\xi^1}.
\end{equation}
Now we can write the vectors in Eq.~(\ref{Frrinpro}) as 
\begin{equation}
\dot{x}_\mu(s') = \tilde{t}_\mu\cosh(as') + \tilde{x}_\mu\sinh(as') = 
\tilde{\eta}_\mu\cosh(a(s-s')) - \tilde{\zeta}_\mu\sinh(a(s-s')),
\end{equation}
\begin{equation}
\frac{x_\mu(s)-x_\mu(s')}{\dot{x}_\nu(s)(x^\nu(s)-x^\nu(s'))} = \tilde{\eta}_\mu \;+\; 
\frac{1-\cosh(a(s-s'))}{\sinh(a(s-s'))}\tilde{\zeta}_\mu,
\end{equation}
Finally, the term responsible for radiation reaction takes the form
\begin{equation}
F_{rr}^\mu = \tilde{\zeta}^\mu\,e^2\int{\drm s'\; \tanh\left(\frac{a}{2}(s-s')\right)\frac{\drm}{\drm s}\Delta^{rr}(x,x')}.
\end{equation}
Since $\Delta^{rr}(x,x')$ ultimately depends on the difference of $s$ and $s'$ we can use 
\linebreak\mbox{$\drm\Delta^{rr}/\drm s = -\drm\Delta^{rr}/\drm s'$} and we get 
\begin{equation}
F_{rr}^\mu = -\tilde{\zeta}^\mu\,e^2\frac{a}{2}\int{\drm s'\;\frac{\Delta^{rr}(x,x')}{\cosh^2\left(\frac{a}{2}(s-s')\right)}}. 
\end{equation}
Using Eq.~(\ref{emab}) we can express $\Delta(x,x')$ as 
\begin{align}
\label{RindProp}
\Delta^{rr}(x,x') &= -\pi i\int{\frac{\ddk}{(2\pi)^4}e^{ik_0(s-s')}\sgn(k_0)\delta(k^2)} \nonumber \\
&= -\frac{1}{4\pi}\sgn(s-s')\delta\left((s-s')^2\right), 
\end{align} 
which allows us to calculate $F_{rr}^\mu$, 
\begin{align}
\label{almostlastFrr}
F_{rr}^\mu &= -\tilde{\zeta}^\mu\,e^2\frac{a}{2}\int{\drm s'\;\frac{\Delta^{rr}(x,x')}{\cosh^2\left(\frac{a}{2}(s-s')\right)}} \nonumber \\ 
&= \tilde{\zeta}^\mu\,a\frac{e^2}{8\pi}\int{\drm s'\;\sgn(s-s')\frac{\delta\left((s-s')^2\right)}{\cosh^2\left(\frac{a}{2}(s-s')\right)}}, \nonumber \\
&= \tilde{\zeta}^\mu\,a\frac{e^2}{8\pi}\int{\drm s'\;\frac{\delta(s-s')}{s-s'}}.
\end{align}
Using Eq.~(\ref{deltatheta}) of appendix B, it can be shown that this result is zero\footnote{Although here it is inconsequential, 
the factor of $\frac{1}{2}$ that would be expected to occur in the last line of Eq.~(\ref{almostlastFrr}) is canceled since $s=s'$ 
is not a simple root of the argument of the Dirac-delta. This is also explained in appendix B where it has more significance.}. Considering that in this frame $\ddot{x}^\mu(s)=0$ and $\dot{x}^\mu(s)=\tilde{\eta}^\mu$, we have 
\begin{equation}
\label{covacc}
\ddot{x}^\mu(s) + \Gamma^\mu_{\alpha\beta}\dot{x}^\alpha(s)\dot{x}^\beta(s) = a\tilde{\zeta}^\mu,
\end{equation}
and thus the equation of motion can be reduced to 
\begin{equation}
m_R a\tilde{\zeta}^\mu = e F^\mu_{e,\,\nu}\tilde{\eta}^\nu. 
\end{equation}

This result can be obtained from Eq.~(\ref{Frrinpro}) with a different approach as well. Knowing that the propagator will 
contribute a Dirac-delta we can expand the expression inside the brackets with respect to $s'$ about $s$, 
\begin{equation}
\dot{x}^\mu(s') - \cosh(a(s-s'))\frac{x^\mu(s)-x^\mu(s')}{\dot{x}_\nu(s)(x^\nu(s)-x^\nu(s'))} = 
\frac{1}{2}\ddot{x}^\mu(s)\, (s'-s) + \mathcal{O}((s'-s)^2).
\end{equation}
Substituting this result into Eq.~(\ref{Frrinpro}) and using $\drm\Delta^{rr}/\drm s = -\drm\Delta^{rr}/\drm s'$ along with 
integration by parts we obtain 
\begin{equation}
F_{rr}^\mu = \frac{e^2}{8\pi} \ddot{x}^\mu(s) \int{\drm s'\;\sgn(s-s')\delta((s-s')^2)}.
\end{equation}
This form of $F_{rr}^\mu$ is still only valid in inertial frames but it can be easily generalized by replacing the acceleration 
four-vector with the covariant expression in Eq.~(\ref{covacc}), 
\begin{equation}
F_{rr}^\mu = \zeta^\mu a\frac{e^2}{8\pi^2} \int{\drm s'\;\frac{\delta(s-s')}{s-s'}}.
\end{equation}
Once again, the result is zero.

This means that the Unruh effect ultimately does not influence the motion of a charged particle that has no excited states. 
If the accelerated charge has multiple levels of internal energy, like the Unruh-DeWitt detector \cite{Unruh1,UnruhWald} or an 
atom \cite{Audretsch:1994yz}, the inertial observer should see Unruh radiation, which is the spontaneous emission of quanta by 
the accelerated object. This claim is also supported by Refs.~\cite{Lima:2018ifz,Zhang:2019}. The corresponding phenomenon in 
the accelerated frame is the excitation of the detector caused by the absorption of quanta from the thermal background. In 
Ref.~\cite{Audretsch:1994yz} it has been shown that the contribution of radiation reaction to the change in energy for both an 
inertial and a uniformly accelerating atom coupled to a scalar field is the same. This implies that radiation reaction vanishes 
even for particles that can be excited by the absorption of photons. From the viewpoint of a co-accelerating observer the 
uniformly accelerated charge should be seen undergoing such an emission process which cancels the effect of absorption. This 
explains why the Unruh effect does not cause recoil and also allows us to check that the existence of the Unruh-effect does 
not violate the equivalence principle. 

We could argue that since the thermal background observed in the accelerated frame is comprised of on-shell excitations as 
evidenced by Eqs.~(\ref{RindlerFeynman}) and (\ref{RindlerWightman}), the charge (which must also be on-shell) should not simply 
absorb photons, but scatter them, like in the case of the Compton effect. Thus, the fact that emission also occurs alongside the 
absorption process should not be surprising. However, considering that the stress-energy tensor satisfies 
$\nabla_\mu T^{\mu\nu}=0$, where $\nabla_\mu$ is the covariant derivative, the sum of kinetic energies, as defined in the 
Rindler frame, is not conserved but there are additional terms involved which take into account that an observer at a fixed 
$\xi^1$ coordinate experiences acceleration $1/\xi^1$. Furthermore, the emission process is such that it cancels the effect of 
absorption exactly. Alternatively, we can describe the phenomenon as photon scattering with zero energy-transfer with a strong 
correlation between the incoming and outgoing momenta of the photon.

\section{A note on bremsstrahlung}
\label{sect:bremsstrahlung}

The world-line of a charge that undergoes hyperbolic motion between proper times $s_1$ and $s_2$ and performs inertial motion 
otherwise, all in the $(x^0,x^1)$ plane, can be written as 
\begin{align}
x^0(s) &= \Theta(s-s_1)\Theta(s_2-s)\frac{1}{a}\sinh(as) \nonumber \\
&\hspace{10pt}+ \Theta(s_1-s)\left[\frac{1}{a}\sinh(as_1)+\cosh(as_1)(s-s_1)\right] \nonumber \\
&\hspace{10pt}+ \Theta(s-s_2)\left[\frac{1}{a}\sinh(as_2)+\cosh(as_2)(s-s_2)\right], \nonumber \\ 
x^1(s) &= \Theta(s-s_1)\Theta(s_2-s)\frac{1}{a}\cosh(as) \nonumber \\
&\hspace{10pt}+ \Theta(s_1-s)\left[\frac{1}{a}\cosh(as_1)+\sinh(as_1)(s-s_1)\right] \nonumber \\
&\hspace{10pt}+ \Theta(s-s_2)\left[\frac{1}{a}\cosh(as_2)+\sinh(as_2)(s-s_2)\right]. 
\end{align}
This correctly yields the acceleration of the charge,  
\begin{equation}
\ddot{x}^\mu(s) = \Theta(s-s_1)\Theta(s_2-s)a\Big(\sinh(as)\tilde{t}^\mu + \cosh(as)\tilde{x}^\mu\Big).
\end{equation}
The ALD radiation reaction for this motion is 
\begin{equation}
F_{ALD}^\mu = \frac{e^2}{6\pi}a\Big[ \delta(s-s_1) - \delta(s-s_2) \Big] \Big(\sinh(as)\tilde{t}^\mu + \cosh(as)\tilde{x}^\mu\Big).
\end{equation}
After the charge resumes inertial motion, the contribution of radiation reaction to the total change in momentum will be 
\begin{align}
\Delta p^\mu_{ALD} &= \frac{e^2}{6\pi} a \Bigg[\Big(\sinh(as_1)-\sinh(as_2)\Big) \tilde{t}^\mu + \Big(\cosh(as_1)-\cosh(as_2)\Big)\tilde{x}^\mu \Bigg] \nonumber \\
&= -\frac{e^2}{6\pi}a^2\Big[\Delta t\,\tilde{t}^\mu + \Delta x\,\tilde{x}^\mu\Big],
\end{align}
where $\Delta t$ and $\Delta x$ are the elapsed coordinate time and the coordinate distance between the start and end of the 
acceleration. We can consider a coordinate frame in which acceleration starts and ends at the same position, i.e.~$\Delta x=0$ and 
it becomes clear that $\Delta p^\mu_{ALD}$ is a time-like vector,  
\begin{equation}
\Delta p^\mu_{ALD} = -\frac{e^2}{6\pi}a^2\Delta t\,\tilde{t}^\mu = - P\Delta t\,\tilde{t}^\mu,
\end{equation}
where the scalar $P>0$ was introduced in Eq.~(\ref{Larmor}). This means that when a charge undergoes uniform acceleration only for 
a finite amount of time there is radiation reaction. In this ideal scenario the contribution of radiation reaction to the change 
in momentum occurs when the acceleration begins and ends but the charge does not experience radiation reaction while it is 
accelerated uniformly. Thus, the results discussed in the previous sections apply even when the charge is accelerated only for a 
finite amount of time and it experiences non-zero radiation reaction only due to transient phenomena.

\section{Radiation reaction through Ehrenfest's theorem}
\label{sect:ehrenfest}

Although in the classical limit the description based on the effective action given by Eq.~(\ref{Seff}) yields the ALD formula for 
radiation reaction, it would be interesting to see whether measurements of a quantum system should yield results that are consistent 
with the ALD theory. Building on the results of section \ref{sect:classical}, we can use Ehrenfest's theorem to see how radiation 
reaction manifests on the level of expectation values. 

If the system starts in state $\phi$ at time $t_1$ and found to be in state $\psi$ at some later time $t_2$, the transition 
amplitude can be written as 
\begin{align}
\langle\psi|\phi\rangle &= \int_{\Sigma_2}{\drm^3x_2\, \int_{\Sigma_1}{\drm^3x_1\, \langle\psi|x_2\rangle \langle x_2|x_1 \rangle \langle x_1|\phi\rangle }} \nonumber \\
&= \int_{\Sigma_2}{\drm^3x_2\, \int_{\Sigma_1}{\drm^3x_1\, \psi^*(x_2)\,\phi(x_1) \langle x_2|x_1 \rangle }},
\end{align}
where $\psi(x_2)$ and $\phi(x_1)$ are the wave functions for states $\psi$ and $\phi$ respectively, integration is carried 
out over appropriate space-like hypersurfaces $\Sigma_1$ at time $t_1$ and $\Sigma_2$ at time $t_2$. The temporal direction is 
chosen to be orthogonal to the hypersurfaces. In this context the coordinates of the particle are observables which can be 
represented by operators. However, as opposed to non-relativistic quantum mechanics, the coordinates of the particle are not 
all independent. 

Now we need to compute how the system evolves between $x_1$ and $x_2$, for which we can use the path integral formalism detailed 
in Ref.~\cite{Feynman}. In this section I repeat Feynman's derivation for a relativistic system in a language consistent with 
the previous sections of this paper. Practically, the system is described in terms of a one-dimensional quantum field theory 
where the so-called field variables are the coordinates of the charge. The evolution of the system is given by 
\begin{equation}
\langle x_2|x_1 \rangle = \int_{\substack{x(s_i)=x_1 \\ x(s_f)=x_2}}{\mathcal{D}x\,\mathcal{D}N\; e^{i\,S[x,N]}},
\end{equation}
where the functional integral is over all paths that start from $x_1$ at initial parameter time $s_i$ and end in $x_2$ at final 
parameter time $s_f$. The dynamics of the system is determined by some action $S[x,N]$ which is a functional of the world-line 
and the lapse function. The integration measure is non-trivial and it is given explicitly in Ref.~\cite{Hartle:1986yu}. 

If we want to compute the transition amplitude (ta) for an arbitrary quantity denoted by $F$ we need to calculate 
\begin{equation}
\langle F\rangle_{ta}\equiv\langle\psi|\hat{F}|\phi\rangle = 
\int_{\Sigma_2}{\drm^3x_2\, \int_{\Sigma_1}{\drm^3x_1\, \psi^*(x_2)\,\phi(x_1) \langle x_2|\hat{F}|x_1 \rangle }},
\end{equation}
where $\hat{F}$ is the operator representation of the quantity over the space of state vectors. In the path integral formalism $F$ 
is represented by a functional $F[x]$ and 
\begin{equation}
\label{transF}
\langle x_2|\hat{F}|x_1 \rangle = \int_{\substack{x(s_i)=x_1 \\ x(s_f)=x_2}}{\mathcal{D}x\,\mathcal{D}N\;F[x]\, e^{i\,S[x,N]}}.
\end{equation}
If the functional $F[x]$ is itself a functional derivative $\delta G[x]/\delta x^\mu(s)$ it follows that 
\begin{equation}
\int_{\substack{x(s_i)=x_1 \\ x(s_f)=x_2}}{\mathcal{D}x\,\mathcal{D}N\;\frac{\delta G[x]}{\delta x^\mu(s)}\, e^{i\,S[x,N]}} 
= - i \int_{\substack{x(s_i)=x_1 \\ x(s_f)=x_2}}{\mathcal{D}x\,\mathcal{D}N\;G[x]\,\frac{\delta S[x,N]}{\delta x^\mu(s)}\, e^{i\,S[x,N]}}.
\end{equation}
Using this we get 
\begin{equation}
\label{GSta}
\left\langle \frac{\delta G}{\delta x^\mu(s)} \right\rangle_{ta} = -i \left\langle G \frac{\delta S}{\delta x^\mu(s)} \right\rangle_{ta}.
\end{equation}
If we substitute unity for G, the expression on the left-hand side will vanish and we are left with 
\begin{equation}
\label{Ehren1}
\left\langle \frac{\delta S}{\delta x^\mu(s)} \right\rangle_{ta} = 0
\end{equation}
which is equivalent to Ehrenfest's theorem.

This formalism, however, cannot accommodate dissipative effects. Instead, we need to describe the effective dynamics of the charge 
using the density operator. The need for this approach is also hinted by the thermal nature of the Unruh effect. The use of path 
integrals for systems interacting with an environment was developed in Ref.~\cite{Feynman:1963118}. 

One possibility is to follow the same logic laid out in the previous part of this section and look for the probability that the 
system starts in a possibly mixed state at $t_1$ characterized by the density operator $\hat{\rho}_i$ and by measurement it is 
found in the state $\psi$ at $t_2$. A similar approach has also been described in Ref.~\cite{Polonyi:2006qc} with the difference 
being in the handling of the final state. The initial density operator can be written in diagonal form as 
\begin{equation}
\hat{\rho}_i = \sum_k{P^{(i)}(\phi_k) |\phi_k\rangle\,\langle\phi_k|},
\end{equation}
where the weight $P^{(i)}(\phi_k)$ represents the probability that the system is in the pure state $\phi_k$ at $t_1$ and 
\begin{equation}
\sum_k{P^{(i)}(\phi_k)} = 1.
\end{equation}
For the final state we need to use the projection operator
\begin{equation}
\hat{\mathtt{P}}_f=|\psi\rangle\langle\psi|
\end{equation}
With this description we can compute only the weighted sum of squared matrix elements but we can make use of the formulae for the 
transition amplitudes from before to obtain expressions built on the path integral formalism, 
\begin{align}
\label{transSME}
\Tr[\hat{\rho}_i\hat{\mathtt{P}}_f] &= \sum_{k}{P^{(i)}(\phi_k) |\langle\psi|\phi_k\rangle|^2} \nonumber \\
&= \sum_{k} P^{(i)}(\phi_k) \int_{\Sigma_1}\drm^3x_1\int_{\Sigma_2}\drm^3x_2 \psi^*(x_2) \phi_k(x_1) 
\int_{\substack{x(s_i)=x_1 \\ x(s_f)=x_2}}{\mathcal{D}x\,\mathcal{D}N\; e^{i\,S[x,N]}} \nonumber \\
&\hspace{55pt} \times\int_{\Sigma_1}\drm^3x'_1\int_{\Sigma_2}\drm^3x'_2 \psi(x'_2) \phi^*_k(x'_1) 
\int_{\substack{x'(s_i)=x'_1 \\ x'(s_f)=x'_2}}{\mathcal{D}x'\,\mathcal{D}N'\; e^{-i\,S[x',N']}},
\end{align}
where $\Tr[...]$ denotes the trace of its argument. Introducing 
\begin{equation}
\rho_i(x_1,x'_1) = \langle x'_1|\hat{\rho}_i|x_1\rangle = \sum_k{P^{(i)}(\phi_k) \phi_k(x_1) \phi^*_k(x'_1)}, 
\end{equation}
we can write Eq.~(\ref{transSME}) as 
\begin{align}
\label{trrhos1}
\Tr[\hat{\rho}_i\hat{\mathtt{P}}_f] = 
\int_{\Sigma_1}\drm^3x_1\int_{\Sigma_2}\drm^3x_2 & \int_{\Sigma_1}\drm^3x'_1\int_{\Sigma_2}\drm^3x'_2\, 
\rho_i(x_1,x'_1)\psi^*(x_2)\psi(x'_2) \nonumber \\
&\times\int{\mathcal{D}x\,\mathcal{D}N\,\mathcal{D}x'\,\mathcal{D}N'\,e^{i\Big(S[x,N]-S[x',N']\Big)}}.
\end{align}
This shows that if we are to calculate the overlap of the density operator at $t_1$ and the state observed at $t_2$ we need to 
use both the forward and backward evolution operator. Namely, the arguments $x_1$ and $x_2$ are connected by forward evolution 
while the arguments $x'_1$ and $x'_2$ are linked together by backward evolution.

In order to interpret the quantity computed in Eq.~(\ref{trrhos1}) let us first discuss the case when we compare states at a fixed 
time. Let us consider a system that is in a mixed state described by the density operator 
\begin{equation}
\hat{\rho}=\sum_i{P(\phi_i) |\phi_i\rangle\langle\phi_i|}, 
\end{equation}
where $P(\phi_i)$ is the probability that the system is in the pure state $\phi_i$ and $|\phi_i\rangle$ are orthonormal. Let 
$\hat{\Lambda}$ be a hermitian operator with eigenvalues $\lambda_i$ and their corresponding eigenvectors $|i\rangle$ that form 
an orthonormal basis over the Hilbert space. Then the pure states $\phi_i$ can be decomposed as 
\begin{equation}
|\phi_i\rangle = \sum_j{c^{(i)}_j |j\rangle}.
\end{equation}
The outcome of a single measurement of the quantity $\Lambda$ will be one of the eigenvalues $\lambda_j$. If the system is in the 
pure state $\phi_i$, the probability of that outcome is given by the conditional probability 
\begin{equation}
P(j|\phi_i) = |\langle j|\phi_i\rangle|^2 = |c^{(i)}_j|^2.
\end{equation}
The trace of the product of the density operator and the projection $|j\rangle\langle j|$ can be written as 
\begin{equation}
\label{rhoprob}
\Tr\Big[\hat{\rho}|j\rangle\langle j|\Big] = \sum_i{P(\phi_i) |\langle j|\phi_i\rangle|^2} = \sum_i{P(\phi_i) P(j|\phi_i)}
= \sum_i{P(j\cap\phi_i)}.
\end{equation}
Furthermore, the sum of all these terms is unity, 
\begin{equation}
\sum_j{\Tr\Big[\hat{\rho}|j\rangle\langle j|\Big]} = \sum_i{P(\phi_i)}\sum_j{|c^{(i)}_j|^2} = 1. 
\end{equation}
Thus the expression in Eq.~(\ref{rhoprob}) can be interpreted as the probability that the outcome of the measurement will be 
$\lambda_j$. The expression in Eq.~(\ref{trrhos1}) is constructed using states at different times and it can be interpreted as 
the transition probability from the possibly mixed state given by $\hat{\rho}_i$ to the pure state $\psi$.

Although Eq.~(\ref{trrhos1}) does not add anything new to the description of systems that are characterized by a simple action 
in the classical limit, we can upgrade the formula by considering instances when there is a difference between the dynamics of 
forward and backward evolution and the exponent of the path integral cannot be separated in the form given above. Such is the 
case when we are interested in radiation reaction. We can write the transition probability as  
\begin{align}
\label{trrhos2}
\Tr[\hat{\rho}_i\hat{\mathtt{P}}_f] = 
\int_{\Sigma_1}\drm^3x_1\int_{\Sigma_2}\drm^3x_2 & \int_{\Sigma_1}\drm^3x'_1\int_{\Sigma_2}\drm^3x'_2\,
\rho_i(x_1,x'_1)\psi(x_2)\psi^*(x'_2) \nonumber \\
&\times\int{\mathcal{D}x\,\mathcal{D}N\,\mathcal{D}\bar{x}\,\mathcal{D}\bar{N}\,e^{i\,S^{CTP}_p[x,N,\bar{x},\bar{N}]}}.
\end{align}
Ordinarily, in the CTP formalism we demand that at the final parameter time $s_f$, every path should end up at the same point, 
i.e.~$x(s_f)=\bar{x}(s_f)$. In Eq.~(\ref{trrhos2}) we need to let go of this constraint.

Given a quantity $F$ we can obtain a weighted sum of squared matrix elements (sme), 
\begin{equation}
\langle F\rangle_{sme} \equiv \Tr[\hat{\rho}_i\hat{F}\hat{\mathtt{P}}_f\hat{F}] = 
\sum_{k}{P^{(i)}(\phi_k) |\langle\phi_k|\hat{F}|\psi\rangle|^2}.  
\end{equation}
When the system can be described by a simple action $S[x,N]$ in the classical limit the weighted sum of squared matrix elements 
can be written in terms of path integrals as 
\begin{align}
\langle F\rangle_{sme} = 
\int_{\Sigma_1}\drm^3x_1\int_{\Sigma_2}\drm^3x_2 & \int_{\Sigma_1}\drm^3x'_1\int_{\Sigma_2}\drm^3x'_2\,
\rho_i(x_1,x'_1)\psi^*(x_2)\psi(x'_2) \nonumber \\
&\times\int{\mathcal{D}x\,\mathcal{D}N\,\mathcal{D}x'\,\mathcal{D}N'\,F[x]F[x'] e^{i\Big(S[x,N]-S[x',N']\Big)}}.
\end{align}
If there are dissipative effects at play, not only we need to use the CTP action, the quantity $F$ might not have a 
representation which is a functional of only one type of path, i.e.~$F[x]$ or $F[x']$. Thus what was written as the weighted sum 
of squared matrix elements before can be computed as 
\begin{align}
\langle F\rangle_{sme} = 
\int_{\Sigma_1}\drm^3x_1\int_{\Sigma_2}\drm^3x_2 & \int_{\Sigma_1}\drm^3x'_1\int_{\Sigma_2}\drm^3x'_2\,
\rho_i(x_1,x'_1)\psi^*(x_2)\psi(x'_2) \nonumber \\
&\times\int{\mathcal{D}x\,\mathcal{D}N\,\mathcal{D}\bar{x}\,\mathcal{D}\bar{N}\,
F[x,\bar{x}] F[\bar{x},x] e^{i\,S^{CTP}_p[x,N,\bar{x},\bar{N}]}}. 
\end{align}

We can take a different road as well, that has been laid out in Refs.~\cite{Feynman:1963118,Polonyi:2012si} and compute the 
expectation value of the quantity $F$ at $t_2$ instead, with only the initial $\hat{\rho}_i$ state of the system known to us. 
The expectation value of $F$ in the final state will be 
\begin{equation}
\langle F \rangle = \int_{\Sigma_1}\drm^3x_1 \int_{\Sigma_1}\drm^3x'_1 \rho_i(x_1,x'_1) 
\int{\mathcal{D}x\,\mathcal{D}N\,\mathcal{D}\bar{x}\,\mathcal{D}\bar{N}\, 
F[x,\bar{x}] e^{i\,S^{CTP}_p[x,N,\bar{x},\bar{N}]}}.
\end{equation}
Following the logic that lead to Eq.~(\ref{Ehren1}), we can use this formula to obtain Ehrenfest's theorem. Assuming that the 
functional $F[x,\bar{x}]$ is itself a functional derivative $\delta G[x,\bar{x}]/\delta x^\mu(s)$ we can write 
\begin{equation}
\left\langle \frac{\delta G}{\delta x^\mu(s)} \right\rangle = 
-i\left\langle G\frac{\delta S^{CTP}_p}{\delta x^\mu(s)} \right\rangle,
\end{equation}
which is analogous to Eq.~(\ref{GSta}) and upon substituting unity for $G$ leads to 
\begin{equation}
\label{ehren2}
\left\langle\frac{\delta S^{CTP}_p}{\delta x^\mu(s)}\right\rangle = 0.
\end{equation}
We can get the same expression for the expectation value of the functional derivatives with respect to $\bar{x}$, $N$ and $\bar{N}$ 
as well. Compared to the classical limit, however, Eq.~(\ref{ehren2}) gives a different result for the expectation values of 
observables. This is due to the fact that in the classical case $x^\mu=\bar{x}^\mu$ but in the quantum mechanical description 
fluctuations can alter the result. If $\langle x^\mu -\bar{x}^\mu \rangle$ remains small, the zeroth-order term in the expansion 
with respect to $x^\mu-\bar{x}^\mu$ yields the same relation between the expectation values of position, velocity, acceleration and 
jerk as Eq.~(\ref{aldprefin}).

Although the discussion of the radiation reaction started from QED, where we compute transition amplitudes between pure states, 
in the derived effective theory we do not have direct access to degrees of freedom associated with the field of the charge since 
integration over them have been carried out. Similarly, in the accelerated frame the Unruh effect can be characterized by a pure 
state but since an observer in the right Rindler wedge cannot gather information about the excitations of the field present in 
the left Rindler wedge, partial tracing must be carried out over some degrees of freedom.

\section{Conclusions}

Classical radiation reaction can be obtained from QED and it is exactly the ALD force. Due to its dissipative nature, radiation 
reaction cannot be discussed within the frames of the ordinary Lagrangian formulation but this problem can be circumvented using 
Schwinger's closed time path method. In the quantum mechanical description the use of density operators is necessitated by the 
dissipative nature of the problem and a relationship between position, velocity, acceleration and jerk analogous to Ehrenfest's 
theorem can be established, although it may differ from the semiclassical prediction when fluctuations become large. The 
reliability of these results is, however, limited. Firstly, it has been assumed that the effect of the charge on the source of 
the external electromagnetic field is negligible and thus the external field does not respond to the charge. Secondly, the 
creation of charged particle--antiparticle pairs and the contribution of charged loops are omitted. 

When a charge undergoes hyperbolic motion, in the accelerated frame it is seen absorbing photons from the thermal background which 
in turn may cause the charge to also emit photons. The net effect of absorption and emission on the motion of the charge is zero. 
This result is consistent with the vanishing radiation reaction obtained in an inertial frame. Thus, in order to observe the 
Unruh effect one must use a detector that has multiple levels of internal energy. However, the Unruh effect does not cause 
recoil while the charge is subjected to uniform acceleration. In the case of a finite-duration hyperbolic motion, there is 
radiation reaction but it contributes to the change in momentum only at the start and the end of acceleration and vanishes 
otherwise. 

Finally, I note that the stimulated emission described in this paper should also occur due to Hawking radiation since near the 
horizon it resembles the Unruh effect \cite{HawkingIsrael}. Further work is required to show whether such an emission is 
directly detectable, however, the fact that the thermal radiation coming from the direction of the event horizon does not 
cause recoil is an important check to verify that the Unruh effect, and by extension the Gibbons-Hawking effect 
\cite{Gibbons:1977mu}, itself does not violate the equivalence principle.

\section*{Appendix A}

In order to obtain the retarded propagator in position space we need to perform partial fraction expansion on its 
Fourier-transform that appears in Eq.~(\ref{propfourier}). Since the propagator in the Feynman gauge is proportional to the 
metric tensor, I drop the Lorentz indices and concentrate only on its scalar coefficient,  
\begin{align}
\Delta^{ret}(x,y)&=\int{ \frac{\ddp}{(2\pi)^4} \frac{e^{-i p\cdot(x-y)}}{(p_0+i0)^2-|\boldsymbol{p}|^2} }
= \int{ \frac{\ddp}{(2\pi)^4} \frac{e^{-i p\cdot(x-y)}}{2|\boldsymbol{p}|} \left[ \frac{1}{p_0-|\boldsymbol{p}|+i0}-\frac{1}{p_0+|\boldsymbol{p}|+i0} \right] } \nonumber \\
&=-\frac{i}{16\pi^3r}\int_{-\infty}^\infty\drm p_0 \int_0^\infty\drm|\boldsymbol{p}|\;e^{-ip_0t}\left(e^{i|\boldsymbol{p}|r}-e^{-i|\boldsymbol{p}|r}\right) \left[ \frac{1}{p_0-|\boldsymbol{p}|+i0}-\frac{1}{p_0+|\boldsymbol{p}|+i0} \right]. 
\end{align}
Here, $t=x^0-y^0$ and $r$ denotes the spatial distance between events at $x^\mu$ and $y^\mu$. The integral with respect to $p_0$ 
can be evaluated using the relations connecting $1/(t-i0)$ and the Heaviside step function $\Theta(\omega)$,  
\begin{align}
\label{HeavisideFourier}
&\int_{-\infty}^\infty{\drm\omega\; \Theta(\pm\omega)e^{-i\omega t}} = \pm\frac{1}{i}\frac{1}{t\mp i0}, 
\\
&\int_{-\infty}^\infty{\drm t\; \frac{1}{t\mp i0} e^{i\omega t}} = \pm 2\pi i \Theta(\pm\omega),
\end{align}
and we obtain the following expression, 
\begin{equation}
\Delta^{ret}(x,y)=-\frac{1}{8\pi^2r}\Theta(t) \int_0^\infty{ \drm|\boldsymbol{p}|\;
\left( e^{-i|\boldsymbol{p}|(t-r)} + e^{i|\boldsymbol{p}|(t-r)} - e^{i|\boldsymbol{p}|(t+r)} - e^{-i|\boldsymbol{p}|(t+r)} \right) }.
\end{equation}
Finally, performing the last integral using Eq.~(\ref{HeavisideFourier}) we get 
\begin{align}
\Delta^{ret}(x,y)&=\frac{i}{8\pi^2r}\Theta(t)\left[ \frac{1}{t-r-i0}-\frac{1}{t-r+i0}-\frac{1}{-t-r-i0}+\frac{1}{-t-r+i0} \right] \nonumber \\
&=-\frac{1}{4\pi r}\Theta(t)\Big[\delta(t-r)-\delta(-t-r)\Big].
\end{align}
Since $r\geq0$, $\Theta(t)\delta(-t-r)=-\Theta(t)\delta(-t-r)=0$ and we can write 
\begin{equation}
\Delta^{ret}(x,y)=-\frac{1}{2\pi}\Theta(t) \delta(t^2-r^2).
\end{equation}
The expression for the advanced propagator can be evaluated similarly, 
\begin{equation}
\Delta^{adv}(x,y)=\int{ \frac{\ddp}{(2\pi)^4} e^{-i p\cdot x} \frac{1}{(p_0-i0)^2-|\boldsymbol{p}|^2} }
=-\frac{1}{2\pi}\Theta(-t)\delta(t^2-r^2).
\end{equation}
and to obtain the Feynman propagator, once again, we need to follow along similar lines, 
\begin{align}
\Delta^F(x,y) &= \int{\frac{\ddp}{(2\pi)^4} e^{-i p\cdot x} \frac{1}{p^2+i0} } 
= \int{ \frac{\ddp}{(2\pi)^4} \frac{e^{-i p\cdot(x-y)}}{2|\boldsymbol{p}|} \left[ \frac{1}{p_0-|\boldsymbol{p}|+i0}-\frac{1}{p_0+|\boldsymbol{p}|-i0} \right] } \nonumber \\
&=\frac{i}{8\pi^2r}\left[ \frac{\Theta(t)}{t-r-i0}+\frac{\Theta(-t)}{t-r+i0}+\frac{\Theta(-t)}{-t-r-i0}+\frac{\Theta(t)}{-t-r+i0} \right] \nonumber \\
&=\frac{i}{8\pi^2r}\left[ \frac{1}{|t|-r-i0} + \frac{1}{-|t|-r+i0} \right] = \frac{i}{4\pi^2} \frac{1}{t^2-r^2-i0}.
\end{align}

Using these results we can write half of the difference of the retarded and advanced propagator as 
\begin{align}
\frac{1}{2}\Big(\Delta^{ret}(x,y)-\Delta^{adv}(x,y)\Big) &= -\frac{1}{4\pi}\sgn(t)\delta(t^2-r^2) \nonumber \\
&= -\pi i\int{ \frac{\ddp}{(2\pi)^4} e^{-i p\cdot(x-y)} \sgn(p_0)\delta(p^2) }, 
\end{align}
and half of the difference of the Feynman propagator and its complex conjugate as
\begin{align}
\frac{1}{2}\Big(\Delta^F(x,y) - \Delta^{F*}(x,y)\Big) &= \frac{i}{4\pi^2}\mathcal{P}\frac{1}{t^2-r^2} \nonumber \\
&= -\pi i\int{ \frac{\ddp}{(2\pi)^4} e^{-i p\cdot(x-y)} \delta(p^2) },
\end{align}
where $\mathcal{P}$ is the Cauchy principal value. This allows us to express the Wightman functions in terms of the four 
solutions listed in Eqs.~(\ref{prop1}) and (\ref{prop2}) as 
\begin{align}
\Delta^{\pm}(x,y)&=-2\pi i\int{ \frac{\ddp}{(2\pi)^4} e^{-i p\cdot x} \Theta(\pm p_0)\delta(p^2) } \nonumber \\
&= \pm\frac{1}{2}\Big(\Delta^{ret}(x,y)-\Delta^{adv}(x,y)\Big) + \frac{1}{2}\Big(\Delta^F(x,y) - \Delta^{F*}(x,y)\Big), 
\end{align}
or alternatively 
\begin{equation}
\Delta^{\pm}(x,y) = \frac{i}{4\pi^2}\frac{1}{(t\mp i0)^2-r^2}.
\end{equation}

\section*{Appendix B}

I will denote the two integrals in Eq.~(\ref{cEOM}) of section \ref{sect:classical} as 
\begin{align}
F_{se}^\mu &= -e^2 \int \drm s'\;\dot{x}^\alpha(s) \dot{x}^\beta(s') 
\Big[ \partial_\alpha\mathrm{Re}\Delta^F_{\mu\beta}(x(s),x(s')) - \partial_\mu\mathrm{Re}\Delta^F_{\alpha\beta}(x(s),x(s')) \Big]
\nonumber \\
F_{rr}^\mu &= e^2 \int \drm s'\;\dot{x}^\alpha(s) \dot{x}^\beta(s') 
\Big[ \partial_\alpha\mathrm{Re}\Delta^+_{\mu\beta}(x(s),x(s')) - \partial_\mu\mathrm{Re}\Delta^+_{\alpha\beta}(x(s),x(s')) \Big]
\end{align}
which are the origins of the self-energy correction and the radiation reaction respectively. Using Eqs.~(\ref{prop2}) and 
(\ref{retadv}) and the Sokhotski-Plemelj theorem it is easily seen that we need to compute the derivatives of the Dirac-delta 
distribution and that of its product with the sign distribution. At this point I introduce the notation $x^\mu\equiv x^\mu(s)$ 
and $y^\mu\equiv x^\mu(s')$. The derivative of the Dirac-delta distribution can be rewritten as 
\begin{align}
\label{deltarel}
\partial_\mu \delta((x-y)^2) = 
2(x_\mu-y_\mu) \delta'((x-y)^2) = 
-\frac{x_\mu-y_\mu}{\dot{y}^\nu(x-y)_\nu}\;\frac{\drm}{\drm s'}\delta((x-y)^2), 
\end{align}
where the prime on the Dirac delta denotes its derivative with respect to its argument. Similarly, the derivative of the product 
of the sign and Dirac delta distributions can be rewritten as 
\begin{equation}
\label{deltarel2}
\partial_\mu \Big[\sgn(x^0-y^0)\delta((x-y)^2)\Big] = 
-\frac{x_\mu-y_\mu}{\dot{y}^\nu(x-y)_\nu}\;\frac{\drm}{\drm s'}\Big[\sgn(x^0-y^0)\delta((x-y)^2)\Big]. 
\end{equation}
The self-energy and radiation reaction terms become 
\begin{align}
F_{se}^\mu &= -\frac{e^2}{4\pi}\dot{x}_\nu\;\int\drm s'\; \Bigg[
\frac{\dot{y}^{[\mu}(x-y)^{\nu]}}{\dot{y}^\alpha(x-y)_\alpha}\; \frac{\drm}{\drm s'}\delta((x-y)^2) 
\Bigg], \nonumber \\
F_{rr}^\mu &= \frac{e^2}{4\pi}\dot{x}_\nu\;\int\drm s'\; \Bigg[
\frac{\dot{y}^{[\mu}(x-y)^{\nu]}}{\dot{y}^\alpha(x-y)_\alpha}\; \frac{\drm}{\drm s'}\Big(\sgn(x^0-y^0)\delta((x-y)^2)\Big) 
\Bigg].
\end{align}
The commutator sign on indices should be interpreted as $A_{[\mu}B_{\nu]}=A_\mu B_\nu - A_\nu B_\mu$. Using integration by parts 
to move the derivative from the expression containing the Dirac delta to the other parts of the integrand we get 
\begin{align}
F_{se}^\mu &= \frac{e^2}{4\pi}\dot{x}_\nu\;\int\drm s'\;\delta((x-y)^2) 
\frac{\drm}{\drm s'}\Bigg[\frac{\dot{y}^{[\mu}(x-y)^{\nu]}}{\dot{y}^\alpha(x-y)_\alpha}\Bigg], \nonumber \\
F_{rr}^\mu &= -\frac{e^2}{4\pi}\dot{x}_\nu\;\int\drm s'\;\sgn(x^0-y^0)\delta((x-y)^2) 
\frac{\drm}{\drm s'}\Bigg[\frac{\dot{y}^{[\mu}(x-y)^{\nu]}}{\dot{y}^\alpha(x-y)_\alpha}\Bigg]. 
\end{align}
Since $x^\mu(s)$ describes the world-line of a massive particle, $(x-y)^2=(x(s)-x(s'))^2$ can be zero only if $s=s'$. However, 
$s'=s$ is not a simple root of the argument. Ordinarily, the Dirac delta of the invariant interval between $x$ and $y$ can be 
given as a combination of two Dirac delta distributions that constrain the integral to retarded and advanced times. When $x$ and 
$y$ are on the same world-line, the retarded and advanced times coincide which results in an additional factor of two, 
\begin{equation}
\delta((x-y)^2) = \frac{1}{|\dot{y}^\beta(x-y)_\beta|}\delta(s-s').
\end{equation}
Furthermore, since the particle is traveling forward in time we have $\sgn(\dot{x}^0)=1$ at every point of the world-line which 
leads to $\sgn(x^0-y^0)=\sgn(s-s')$. Thus the self-energy and radiation reaction terms will be 
\begin{align}
F_{se}&=\frac{e^2}{4\pi}\dot{x}_\nu\;\int\drm s'\; \frac{\delta(s-s')}{|\dot{y}^\beta(x-y)_\beta|} 
\frac{\drm}{\drm s'}\Bigg[\frac{\dot{y}^{[\mu}(x-y)^{\nu]}}{\dot{y}^\alpha(x-y)_\alpha}\Bigg], \nonumber \\
F_{rr}&=-\frac{e^2}{4\pi}\dot{x}_\nu\;\int\drm s'\; \frac{\delta(s-s')}{|\dot{y}^\beta(x-y)_\beta|} \Bigg\{
\sgn(s-s')\frac{\drm}{\drm s'}\Bigg[\frac{\dot{y}^{[\mu}(x-y)^{\nu]}}{\dot{y}^\alpha(x-y)_\alpha}\Bigg] \Bigg\}.
\end{align}
Now, let us expand $y^\mu\equiv x^\mu(s')$ in $s'$ about $s$. In order to make the following expressions easier to handle, I 
will denote $\dels=s'-s$. Due to the presence of the Dirac delta in both integrals, all $\mathcal{O}(\dels)$ terms of both 
integrands will vanish and only the pole and the finite part will remain. Thus, as we will see, it is sufficient to expand 
$y_\mu$ up to third order in $s'$, 
\begin{align}
y_\mu &= x_\mu + \dot{x}_\mu\dels + \frac{1}{2}\ddot{x}_\mu\dels^2 + \frac{1}{6}\dddot{x}_\mu\dels^3 + \mathcal{O}(\dels^4), \nonumber \\
\dot{y}_\mu &= \dot{x}_\mu + \ddot{x}_\mu\dels + \frac{1}{2}\dddot{x}_\mu\dels^2 + \mathcal{O}(\dels^3), \nonumber \\
\dot{y}^\mu(y-x)_\mu &= \dels + \left(\frac{2}{3}\dot{x}^\mu\dddot{x}_\mu + \frac{1}{2}\ddot{x}^\mu\ddot{x}_\mu\right)\dels^3 + \mathcal{O}(\dels^4), \nonumber \\
\dot{y}_{[\mu}(y-x)_{\nu]} &= -\frac{1}{2}\dot{x}_{[\mu}\ddot{x}_{\nu]}\dels^2 + \frac{1}{3}\dddot{x}_{[\mu}\dot{x}_{\nu]}\dels^3 + \mathcal{O}(\dels^4)
\end{align}
Combining these expressions we get 
\begin{equation}
\frac{\dot{y}_{[\mu}(x-y)_{\nu]}}{\dot{y}^\alpha(x-y)_\alpha} = -\frac{1}{2}\dot{x}_{[\mu}\ddot{x}_{\nu]}\dels + \frac{1}{3}\dddot{x}_{[\mu}\dot{x}_{\nu]}\dels^2 + \mathcal{O}(\dels^3).
\end{equation}
Now the integrand of $F_{se}^\mu$ without the Dirac delta becomes
\begin{equation}
\label{smallpart1}
\frac{1}{|\dot{y}^\beta(x-y)_\beta|}\frac{\drm}{\drm s'}\left[\frac{\dot{y}_{[\mu}(x-y)_{\nu]}}{\dot{y}^\alpha(x-y)_\alpha}\right] = 
-\frac{1}{2}\dot{x}_{[\mu}\ddot{x}_{\nu]}\frac{1}{|\dels|} + \frac{2}{3}\dddot{x}_{[\mu}\dot{x}_{\nu]}\sgn(s'-s) + \mathcal{O}(\dels), 
\end{equation}
while the integrand of $F_{rr}^\mu$ without the Dirac delta turns into  
\begin{equation}
\label{smallpart2}
\frac{\sgn(s-s')}{|\dot{y}^\beta(x-y)_\beta|}\frac{\drm}{\drm s'}\left[\frac{\dot{y}_{[\mu}(x-y)_{\nu]}}{\dot{y}^\alpha(x-y)_\alpha}\right] = 
\frac{1}{2}\dot{x}_{[\mu}\ddot{x}_{\nu]}\frac{1}{\dels} - \frac{2}{3}\dddot{x}_{[\mu}\dot{x}_{\nu]} + \mathcal{O}(\dels). 
\end{equation}
Note that the sign of $\dels$ is the opposite of $\sgn(s-s')$. Both expressions are multiplied by $\delta(\dels)$ and integrated 
with respect to $s'$ in the equation of motion. The $\mathcal{O}(\dels)$ terms will not give any contribution in either term. 
Furthermore, the second term of Eq.~(\ref{smallpart1}) and the first term of Eq.~(\ref{smallpart2}) also vanishes. This can be 
shown by using
\begin{equation}
\label{deltatheta}
\delta(x) = \lim_{\epsilon\rightarrow0+}\frac{1}{\epsilon}\Theta\left(\frac{\epsilon}{2}-|x|\right).
\end{equation}
Thus, 
\begin{equation}
\int{\drm s'\;\sgn(s'-s)\delta(s-s')} = 
\lim_{\epsilon\rightarrow0+}\frac{1}{\epsilon}\int_{-\epsilon/2}^{\epsilon/2}{\drm\dels \sgn(\dels)} = 0, 
\end{equation}
and 
\begin{equation}
\int{\drm s'\; \frac{\delta(s-s')}{s-s'}} = 
\lim_{\epsilon\rightarrow0+}\frac{1}{\epsilon}\int_{-\epsilon/2}^{\epsilon/2}{\frac{\drm\dels}{\dels}} = 
\lim_{\epsilon\rightarrow0+}\frac{1}{\epsilon}\lim_{\delta\rightarrow0+}\Bigg[
\int_{\delta}^{\epsilon/2}{\frac{\drm\dels}{\dels}}+\int_{-\epsilon/2}^{-\delta}{\frac{\drm\dels}{\dels}}
\Bigg] = 0. 
\end{equation}
The first term of Eq.~(\ref{smallpart1}) gives a divergent contribution for the self-energy 
\begin{align}
F_{se}^\mu &= \frac{e^2}{8\pi}\ddot{x}^\mu\int_{-\infty}^\infty{\drm s'\,\frac{\delta(s'-s)}{|s'-s|}}
=\frac{e^2}{8\pi}\ddot{x}^\mu\int_{-\infty}^\infty{\drm s'\,\frac{1}{|s'-s|}\frac{\drm\Theta(s'-s)}{\drm s'}} \nonumber \\
&=\frac{e^2}{8\pi}\ddot{x}^\mu\int_s^\infty{\drm s'\,\frac{1}{(s'-s)^2}}
=\frac{e^2}{8\pi}\ddot{x}^\mu\lim_{\epsilon\rightarrow0+}\frac{1}{\epsilon}, 
\end{align}
while second term of Eq.~(\ref{smallpart2}) gives a finite contribution. Using these results in 
the equation of motion we obtain 
\begin{equation}
m\ddot{x}^\mu - e F_{e,\,\nu}^\mu\dot{x}^\nu - \frac{e^2}{8\pi}\Bigg(\lim_{\epsilon\rightarrow0+}\frac{1}{\epsilon}\Bigg)\ddot{x}^\mu
- \frac{e^2}{6\pi}\Big[\dddot{x}^\mu - \dot{x}^\mu (\dddot{x}^\nu\dot{x}_\nu)\Big] = 0.
\end{equation}
The divergent expression can be removed by defining
\begin{equation}
m_R = m - \frac{e^2}{8\pi}\lim_{\epsilon\rightarrow0+}\frac{1}{\epsilon}, 
\end{equation}
and the equation motion can be put in the form 
\begin{equation}
m_R\ddot{x}^\mu - e F_{e,\,\nu}^\mu\dot{x}^\nu 
- \frac{e^2}{6\pi}\Big[\dddot{x}^\mu - \dot{x}^\mu (\dddot{x}^\nu\dot{x}_\nu)\Big] = 0.
\end{equation}


\bibliographystyle{ieeetr}
\bibliography{ALDfromQED_rev3}

\end{document}